\documentclass[10pt,reqno,oneside]{amsart}
\usepackage{cancel,ulem}
\usepackage{amsmath,amsfonts,amssymb,amsthm,subfig,verbatim,multicol,color,verbatim}
\usepackage{graphicx}
\usepackage{natbib}
\usepackage{hyperref}

\numberwithin{equation}{section}

\theoremstyle{definition}

\theoremstyle{remark}

\newcommand{\RR}{\mathbb{R}}      
\newcommand{\Real}{\mathbb{R}}
\newcommand{\ZZ}{\mathbb{Z}}      

\newcommand{\curl}{\operatorname{curl}}

\numberwithin{equation}{section}

\begin{document}

\title[Vortex Gas Models for Tornadogenesis and Maintenance]{Applications of Vortex Gas Models to Tornadogenesis and Maintenance}

\author{Pavel B\v{e}l\'{\i}k}
\author{Douglas P.~Dokken}
\author{Corey K.~Potvin}
\author{Kurt Scholz}
\author{Mikhail M.~Shvartsman}

\address{Pavel B\v{e}l\'{\i}k\\
Department of Mathematics, Statistics, and Computer Science\\
Augsburg University\\
2211 Riverside Avenue\\
Minneapolis, MN 55454\\
U.S.A.} \email{belik@augsburg.edu}

\address{Douglas P.~Dokken\\
Department of Mathematics\\
University of St.~Thomas\\
2115 Summit Avenue\\
St.~Paul, MN 55105\\
U.S.A.} \email{dpdokken@stthomas.edu}

\address{Corey K.~Potvin\\
NSSL/FRDD Rm 4378\\
120 David L.~Boren Boulevard\\
Norman, OK 73072\\
U.S.A.} \email{corey.potvin@noaa.gov}

\address{Kurt Scholz\\
Department of Mathematics\\
University of St.~Thomas\\
2115 Summit Avenue\\
St.~Paul, MN 55105\\
U.S.A.} \email{k9scholz@stthomas.edu}

\address{Mikhail M.~Shvartsman\\
Department of Mathematics\\
University of St.~Thomas\\
2115 Summit Avenue\\
St. Paul, MN 55105\\
U.S.A.} \email{mmshvartsman@stthomas.edu}

\thanks{The four authors, B\v{e}l\'{\i}k, Dokken, Scholz, and Shvartsman were supported by National Science Foundation grant DMS-0802959. Funding for Potvin was provided by the NOAA/Office of Oceanic and Atmospheric Research under NOAA--University of Oklahoma Cooperative Agreement \#NA11OAR4320072, U.S.~Department of Commerce. 
}



\date{\today}

\begin{abstract}
  Processes related to the production of vorticity in the forward and rear flank downdrafts and their interaction with the boundary layer are thought to play a role in tornadogenesis. We argue that an {\it inverse energy cascade} is a plausible mechanism for tornadogenesis and tornado maintenance and provide supporting evidence which is both numerical and observational. We apply a three-dimensional vortex gas model to supercritical vortices produced at the surface boundary layer possibly due to interactions of vortices brought to the surface by the rear flank downdraft and also to those related to the forward flank downdraft. Two-dimensional and three-dimensional vortex gas models are discussed, and the three-dimensional vortex gas model of Chorin, developed further by Flandoli and Gubinelli, is proposed as a model for intense small-scale subvortices found in tornadoes and in recent numerical studies by Orf et al. In this paper, the smaller scales are represented by intense, supercritical vortices, which transfer energy to the larger-scale tornadic flows (inverse energy cascade). We address the formation of these vortices as a result of the interaction of the flow with the surface and a boundary layer.
\end{abstract}

\allowdisplaybreaks
\thispagestyle{empty}
\maketitle

\section{Introduction}
\label{sec:introduction}
In classical statistical mechanics and thermodynamics, one attempts to explain the macroscopic behavior of gases by using the statistics of modeled microscopic behavior of the individual gas molecules and their interactions. In analogy with this theory, the interaction of large numbers of vortices in two- and three-dimensional space has been studied by modeling the vortices as part of a vortex gas. This theory has its origins in the $19$th century in the works of \citet{helmholtz58} and \citet{kelvin69}. \citet{onsager49} first introduced the notion of entropy and temperature for vortex gases that is different from the usual notions of entropy and temperature of gases of molecules, and formulated a two-dimensional theory. In the statistical mechanics context, negative temperatures are higher than positive temperatures. In the two-dimensional vortex gas case, the molecules are replaced by point vortices; in the three-dimensional case, they could be arching vortex lines (tubes) or segments of a single vortex (in a collection of several vortices). In both these cases, negative temperatures are conceivable. The three-dimensional vortex gas model of \citet{chorin-akao91,chorin91,chorin}, developed further by \cite{flandoligubinelli02}, can be applied to model the behavior of intense three-dimensional vortices anchored at the surface, and thus it can contribute to the understanding of the processes of tornadogenesis and maintenance.

It is generally accepted that the process of tornadogenesis involves vortices generated baroclinically by the rear flank downdraft that are then turned into the vertical (see, e.g., \citet{naylorgilmore13,markowskirichardsonbryan14}) and then potentially anchored at the surface. Recent numerical simulations of \citet{orfwilhelmsonwickerleefinley14,orf16} indicate two potentially important factors contributing to tornadogenesis: (a) consolidation of vertical vortices generated along the forward flank downdraft boundary and anchored at the surface that enter and strengthen the developing tornado; and (b) a streamwise current of horizontally generated vorticity that is tilted upward into the low-level mesocyclone. \cite{sasaki14} proposed a theory of the balance of thermodynamic entropy to explain the process of the generation of vorticity in the parent supercell storm, the related development of rotation at the surface, and the subsequent formation and maintenance of a tornado; as the surface tornado development takes place, the vorticity becomes dominated by barotropic vorticity.

In this paper we attempt to provide a more detailed study of the development of rotation at the surface and the subsequent tornadogenesis and maintenance. The main idea in this paper is the inverse energy cascade supported by vortex gas models, whereby the energy from small-scale, intense vortices is transferred to the large-scale tornadic flow. This process would support tornadogenesis and tornado maintenance. \citet{lewellensheng80} argue that these intense vortices within the tornado circulation can be interpreted as turbulent eddies. As a consequence of this study, \citet{wilhelmsonwicker02} noted that the turbulent eddies represented a significant portion of the kinetic energy of the flow. In a modeling study, \citet{fiedler94} found that such steady vortices can produce wind-speeds roughly twice the thermodynamic speed limit, while non-steady transient vortices can produce velocities six times the thermodynamic speed limit. In the context of three-dimensional turbulence, \citet{chorin-akao91,chorin91,chorin} argue that intense vortices have negative temperature. A vortex is said to have negative temperature in the statistical mechanics sense if an increase in energy of the system results in a decrease in its entropy. We draw a parallel between the large eddies used to represent intense vortices in tornadoes and negative-temperature vortices in turbulent flows. That is, we argue that the small-scale, intense vortices in tornadic flows have negative temperature, presumably higher than that of the ambient vortex; consequently, they transfer energy to the surrounding flow and at the same time increase their own entropy. In this process, these intense vortices fold and kink up tightly, and dissipate. The result is the intensification of the surrounding tornadic flow, thus potentially leading to tornadogenesis or contributing to the energy maintenance or intensification of the existing tornado. Similar observations have been made in the early simulations of \citet{orfwilhelmsonwickerleefinley14,orf16}. Also, recent work of \citet{nolan12} suggests that a significant amount of the perturbation energy in tornadoes is due to stretching of asymmetries by the updraft near the surface, ``bringing the flow closer to solid body rotation,'' further supporting the idea of an inverse energy cascade.

In \citet{larchevequechaskalovic94} bifurcation theory is applied to a nonrotating updraft: they perturb the updraft with a rotating eddy and the updraft acquires rotation, the vorticity generated by the perturbations is focused, and a vortex forms. The model they use is based on Serrin's swirling vortex. They apply bifurcation theory developed by \cite{temam77} to study the Taylor--Couette flow. This appears to be similar to the process observed in the numerical simulations of \citet{orfwilhelmsonwickerleefinley14,orf16}, whereby tornadogenesis is initiated by the infusion of a sequence of vortices (eddies) from the forward flank region of the storm into the region below the mesocyclone updraft.


Our general framework is to use a Lagrange multiplier argument to maximize entropy of a collection of vortices or a single, isolated vortex. The result is, among other things, a temperature of such a system, the maximum entropy, and other relevant Lagrange multiplier(s). A system in equilibrium has maximum entropy. Locally in time a system would adjust to equilibrium (subject to the constraints). When a system interacts with another system, the entropy of the combined system would adjust. We make an assumption of finite energy levels (or a bounded range) which is reasonable for a bounded system. This allows the possibility of negative temperatures for a vortex system and a single vortex in three dimensions. We believe that the process of adjustment to maximum entropy and the notion of negative temperature apply to tornadoes in general.

The process we describe may augment and clarify other tornado-related processes. This process is highly discrete and not microscopic. The tracks of the suction vortices (supercritical vortices) in a tornado reveal that the number of vortices is not very large. One could speculate that there are a large number of such vortices at a microscopic scale (analogous to the carbonated-water tornado vortex model, see \citet{turnerlilly63}), but we are not doing that. The supercritical vortices are not really microscopic, though they are relatively small and have a high energy density. The vortex gas theory is about the raising of or maintaining the temperature of the tornado through an interaction with higher-temperature supercritical vortices. The image of many very small microscopic vortices bombarding the tornado is not what we are trying to convey, but rather it is a more discrete and slower process modeling the interaction of discrete suction vortices with a larger tornadic or pre-tornadic flow.

\begin{figure*}
  \begin{center}
    \includegraphics[height=0.93\textheight]{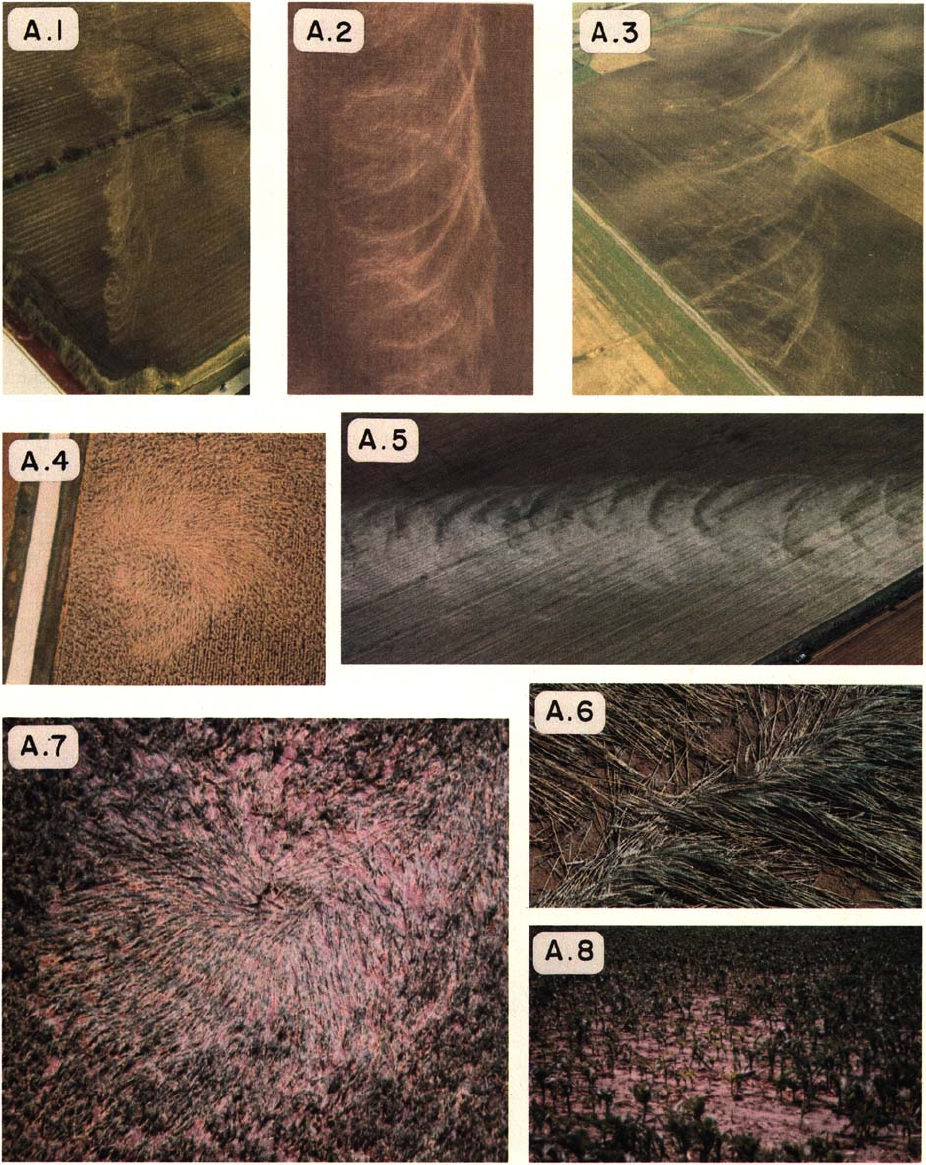}
  \end{center}
  \caption{Tracks left in corn fields showing vortices spiraling into tornadoes and then dissipating; \copyright~AMS, \citet{fujita81}.}
  \label{fig:suctionspots}
\end{figure*}

Besides the computational evidence, there is also empirical evidence of the existence of such high-intensity vortices. Some are shown in Figure \ref{fig:suctionspots} and discussed in \citet{fujita81}, for example.
Tracks left behind by such vortices within a tornado could be as narrow as $30$ cm, too small to be detected with current radar technology. Some of these paths appear to originate outside the tornado and intensify as they move into the tornado. We identify these vortices as supercritical in the sense of \citet{fiedlerrotunno86}, and based on their analysis and the work of \citet{barcilon67,burggraf77,benjamin62} we will argue that they have negative temperature. This will be based on the observation that as the angular momentum increases, the energy density of the supercritical vortex increases, and its entropy density, viewed as a randomness in the vortex, decreases.

In this paper, we review many theoretical concepts from statistical mechanics that have been applied to fluids and interpret them in the context of tornadoes. In particular, we discuss the concepts of negative temperature, two- and three-dimensional vortex gas models, and the inverse energy cascade. We also suggest ways in which these concepts help to explain tornadogenesis and maintenance.

We organize the paper as follows. In section \ref{sec:vortexstretching} we discuss the background fluid mechanics and the notion of ``supercritical'' suction vortices. We briefly discuss their characteristics and origin, and provide some heuristic arguments for the later sections.

In section \ref{sec:statmech} we introduce a statistical mechanics approach to describe vortex gases, which is applicable in both two and three dimensions. Our primary goal is to use this idea in three dimensions. We note that we are not using any two-dimensional statistical mechanics assumptions and that the argument works for three-dimensional vortices. To achieve the generality needed for the rest of the paper, we consider a system in which microstates can be associated with a finite number (or range) of levels of relevant physical macroscopic quantities, and discuss a Lagrange multipliers approach that allows one to introduce the notion of negative temperature in the system.

In section \ref{sec:2Dvortexgas} we first introduce the two-dimensional vortex gas model with ``particles'' of the system being point vortices in the plane, and discuss the behavior of some such systems, both with positive and negative temperature.

In section \ref{sec:3Dvortexgas} we review several models for vortex gases in three dimensions. We first discuss models for nearly parallel vortices and then focus on models that allow vortices to stretch, fold, and dissipate. This notion will be important in the context of the behavior of supercritical vortices in larger tornadic flows, which is discussed in section \ref{sec:entropy-and-temperature}. In section \ref{sec:entropy-and-temperature} we also discuss the notion of entropy, energy, and temperature of interacting systems, starting from a classical point of view of \citet{landaulifshitz}. We discuss the similarities between supercritical vortices in tornado vortex chambers and in nature, and argue that the supercritical vortices may provide a mechanism for both the maintenance as well as the genesis of a tornado.

Conclusions and further discussion are presented in section \ref{sec:conclusions}.

\section{Background and suction vortices}
\label{sec:vortexstretching}
Fluid flows with large Reynolds numbers are modeled using Euler's equations that relate the velocity, pressure, and density of the fluid and omit the negligible viscosity effects. A typical assumption is that the flow is incompressible, i.e., the velocity field is divergence free. It can be shown theoretically that an isentropic flow is nearly incompressible if the flow speeds, or the local changes in flow speeds, along streamlines are small compared to the speed of sound of the medium (\citet{chorin-marsden93}). Numerical simulations of \citet{xialewellens03} show agreement of results for intense tornadic compressible and incompressible isentropic flows. In what follows, we will thus assume the flow is incompressible.

The governing equations for an incompressible fluid flow are
\begin{gather*}
  \frac{D{\bf u}}{Dt}
  =
  \frac{\partial{\bf u}}{\partial t}
  +  ({\bf u}\cdot\nabla){\bf u}
  =
  -\frac{1}{\rho}\nabla p+{\bf b},\\
  \nabla\cdot{\bf u}=0,\\
  \dfrac{D\rho}{Dt}=0,
\end{gather*}
where $\bf u$ is the fluid's velocity, $\rho$ is the fluid's mass density, $p$ is its pressure, and $\bf b$ is an external body force.

Let $\boldsymbol\xi$ denote the vorticity of $\bf u$, i.e., ${\boldsymbol\xi}=\curl{\bf u}=\nabla\times{\bf u}$. Assuming a conservative body force, $\nabla\times\bf b=0$, one can obtain an equation for $\boldsymbol\xi$ (vorticity equation),
\begin{equation}
  \label{eq:DomegaDt}
  \frac{\partial\boldsymbol\xi}{\partial t}
  =
  \nabla\times({\bf u}\times\boldsymbol\xi)
  +
  \frac{1}{\rho^2}\nabla\rho\times\nabla p.
\end{equation}
In this equation, the first term on the right corresponds to the ``barotropic'' generation of vorticity (and captures the advection, stretching, and tilting of the vertical vorticity; see, e.g., \citet{klemp87}), while the second term on the right corresponds to the ``baroclinic'' generation of vorticity, i.e., vorticity generation due to the misalignment of the gradients of mass density and pressure.

In the rest of the paper, we will focus on flows in which a significant amount of vorticity is supported on long, narrow vortex filaments, which may be embedded in a larger rotational flow. These may be baroclinic or barotropic in origin, and may demonstrate themselves, e.g., as the suction spots shown in Figure \ref{fig:suctionspots}.

Flows with vortices that form in strongly sheared environments have a roughly two-dimensional structure before they are stretched. In particular, \citet{pouquet10} note that ``under the influence of a strong external agent, such as gravity, rotation or magnetic fields [\dots] the flow becomes anisotropic and, in fact, tends to (although it never reaches) a near two-dimensional state, with thin layers in the case of stratification, or columnar (Taylor) vortices in the case of rotation.'' We will argue later, based on earlier work of \citet{chorin-akao91,chorin}, that three-dimensional effects have to be taken into consideration when such vortices are stretched, due to their kinking up and dissipation.

Some of the tracks shown in Figure \ref{fig:suctionspots} exhibit similarities to the two-dimensional case of interacting point vortices (see, e.g., \citet{lim-nebus07} and section \ref{sec:2Dvortexgas} below). The implication is that suction vortices appear to behave like two-dimensional vortices, then dissipate, potentially due to stretching. This phenomenon can be observed in videos of intense tornadoes and appears to be present in the simulations of \citet{orfwilhelmsonwickerleefinley14,orf16}. Related concepts are discussed by \citet{fiedlerrotunno86,fiedler94,fiedler97,xialewellens03,lewellens07}.

These high-intensity vortices are barotropic, however, their origin could very well be baroclinic. Recent results of \citet{markowskirichardsonbryan14} suggest vorticity is produced baroclinicly in the rear flank downdraft, it then descends to the surface and is tilted into the vertical, and this process is linked to tornadogenesis. Once these vortices come into contact with the surface, and the stretching and surface friction related swirl (boundary layer effects) are in the appropriate ratio, then by analogy with the work of \citet{fiedlerrotunno86} discussed below the vortex would now be barotropic and could have ``negative temperature.''

\begin{figure*}
  \begin{center}
  \includegraphics[height=0.29\textheight]{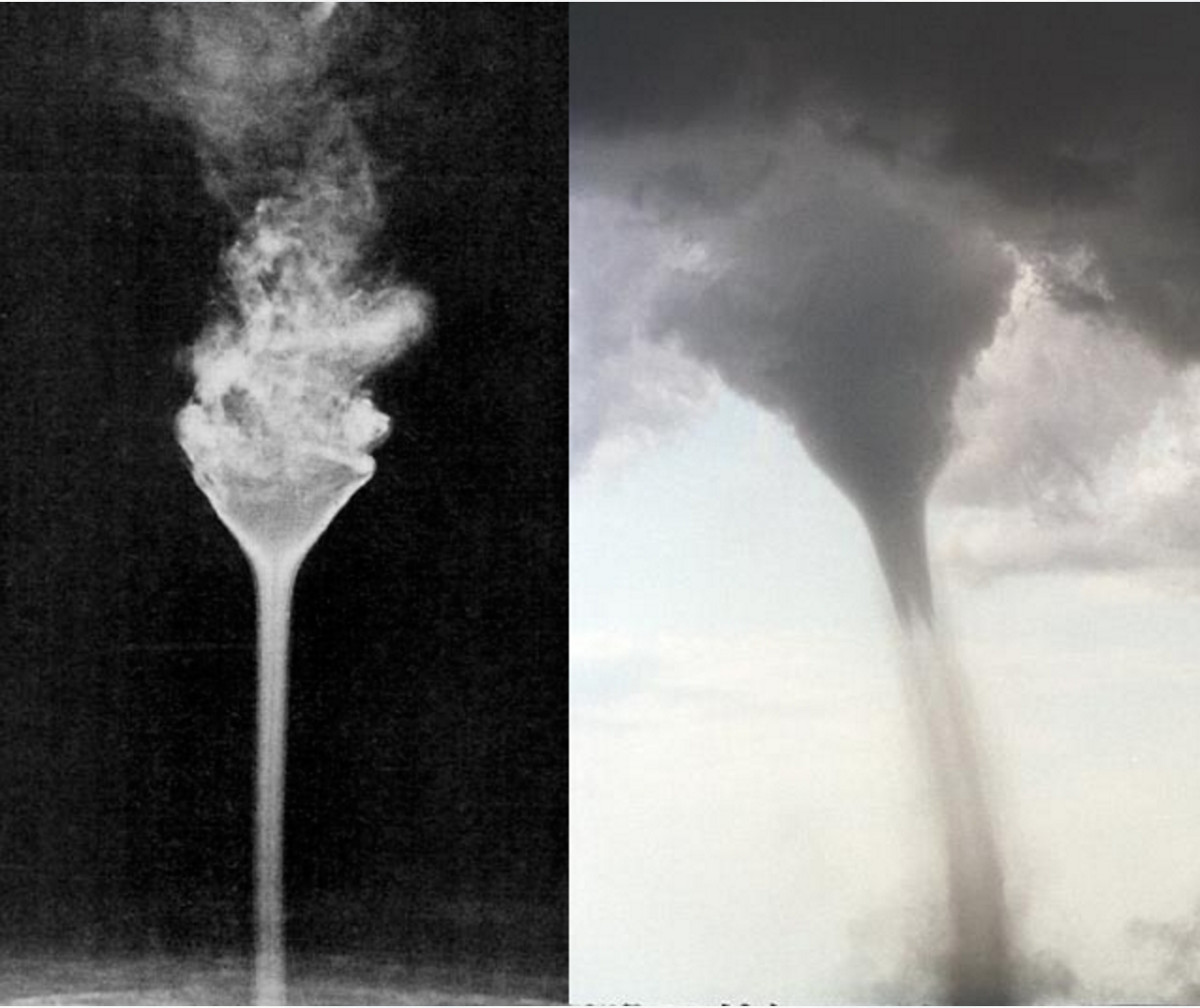}
  \end{center}
  \caption{Vortex breakdown midway through the vortex. Borrowed from \url{http://qdl.scs-inc.us/2ndParty/Pages/7199.html}.}
  \label{fig:champagne}
\end{figure*}

In statistical mechanics, a physical system is said to have negative temperature if, when energy is added to the system, the entropy of the system decreases, i.e., the system becomes ``less random.'' The supercritical vortex below a breakdown bubble is an example of such a system, as follows from the following discussion. Studies of vortex behavior in a Ward chamber (\citet{ward72,davies-jones73,church77}) have shown that when the swirl ratio is in a certain range, the flow configuration takes on a structure that resembles a champagne glass. The stem of the glass would correspond to the supercritical vortex and the part of the glass above the stem to the breakdown bubble
(see Figure \ref{fig:champagne}). Theoretical studies by \citet{barcilon67,burggrafstewartsonbelcher71,burggraf77}, and \citet{fiedlerrotunno86}, and experimental studies by \citet{ward72} and later \citet{church77} contributed to the identification of relationships between various quantities of interest in a supercritical vortex.

Experimentally the following has been found for the vortex chamber flows. Holding the upward volumetric flow constant, as angular momentum is increased, the radius of the supercritical vortex decreases, as does the length of the supercritical vortex, while the axial and azimuthal components of the velocity increase. This suggests that when the angular momentum is increased, the vortex is becoming ``less random'' as its volume decreases (entropy density is decreasing) and the volumetric energy density of the vortex is increasing as the velocity increases. Hence such a vortex would exhibit negative temperature.

As argued by \citet{fiedlerrotunno86}, it is not unreasonable to expect behavior similar to that observed in vortex chambers also in tornadic vortex flows. The critical aspect of the argument is that the vortices must be in contact with the ground for the supercritical, negative-temperature vortex to occur. The effects of the boundary layer are critical in that as swirl increases, the thickness of the boundary layer decreases, and the radius of the supercritical vortex (thought of as an extension of the boundary layer) also decreases, as well as the length of the vortex. Additionally, as swirl increases, the axial velocity in the core and the azimuthal velocity increase, suggesting that the energy density of the vortex increases.

In the next sections, we will introduce the statistical mechanics of vortex gases and then discuss the two- and three-dimensional point vortex theories. The simpler two-dimensional theory allows one to gain some insights prior to any of the three-dimensional effects becoming important.

\section{Statistical mechanics of vortex gases}
\label{sec:statmech}
We next give a brief exposition of a theory of vortex gases by proceeding in analogy with the development of the Boltzmann distribution in the theory of statistical mechanics.

The theory we describe below can be applied in both two and three dimensions and can take different forms. Specifically, we will assume that a certain range of energy levels is accessible to the vortex system and study the energy probability distribution of the system as a whole, given, for example, a mean energy of the ensemble. This setting does not require a particular large number of vortices.

As a specific example in two dimensions, one can replace the particles by point vortices and study the statistics of their distribution in the corresponding phase space with various conserved quantities serving as constraints in the Lagrange multipliers argument presented below. A specific example in three dimensions will be described in section \ref{sec:3Dvortexgas}.

Consider a system in which microstates can be associated with only a finite number of levels of relevant physical quantities. For the sake of simplicity, but to illustrate the argument in a fairly general form, we will describe below the case with energy and another quantity, for example, a moment of inertia, being the conserved quantities. One can easily modify this argument to use fewer or more constraints as seen below.

We consider a vortex system with $k$ possible energy and moments of inertia levels $(E_j,I_j)$, $j=1,\ldots,k$. Let $0\le p_j\le1$ represent the probability that the system is in a state with energy $E_j$ and moment of inertia $I_j$. We assume that the macroscopic quantities of the system, the mean energy of the ensemble, denoted by $\langle E\rangle$, and the mean moment of inertia, denoted by $\langle I\rangle$, are fixed. Then
\begin{equation}
  \label{eq:constr1}
  \sum_{j=1}^k p_j
  =
  1,
\end{equation}
\begin{equation}
  \label{constr2}
  \sum_{j=1}^k p_j E_j
  =
  \langle E\rangle,
\end{equation}
\begin{equation}
  \label{constr3}
  \sum_{j=1}^k p_j I_j
  =
  \langle I\rangle.
\end{equation}
We define the entropy of the ensemble corresponding to the macrostate $(\langle E\rangle,\langle I\rangle)$, up to an additive constant, as
\begin{equation}
  \label{eq:entropy}
  S
  =
  -\sum_{j=1}^k p_j\log{p_j}.
\end{equation}
To maximize the entropy, we consider the Lagrangian
\begin{equation}
  \label{unconstrained}
    L(p_1,\ldots,p_k)
    =
    -
    \sum_{j=1}^k p_j\log{p_j}
    -
    \alpha\left(\sum_{j=1}^k p_j-1\right)
    -
    \beta\left(\sum_{j=1}^k p_j E_j-\langle E\rangle\right)
    -
    \gamma\left(\sum_{j=1}^k p_j I_j-\langle I\rangle\right),
\end{equation}
where $\alpha$, $\beta$ and $\gamma$ are Lagrange multipliers. Differentiating \eqref{unconstrained} with respect to each of the $p_j$ and setting these partial derivatives equal to zero, we obtain
\begin{equation*}
  -\log{p_j}-1-\alpha-\beta E_j-\gamma I_j
  =
  0,
  \qquad
  j=1,\dots,k,
\end{equation*}
while the derivatives with respect to the Lagrange multipliers return the constraints \eqref{eq:constr1}--\eqref{constr3}. This results in
\begin{equation}
  \label{pj}
  p_j
  =
  e^{-1-\alpha-\beta E_j-\gamma I_j},
  \qquad
  j=1,\dots,k.
\end{equation}
From \eqref{eq:constr1} and \eqref{pj} we now have
\begin{equation}
  \label{eq:partition_function}
  e^{1+\alpha}
  =
  \sum_{j=1}^k e^{-\beta E_j-\gamma I_j}
  \equiv
  Z,
\end{equation}
where $Z$ is called a partition function. It follows that
\begin{equation}
  \label{eq:p_j}
  p_j
  =
  \frac{e^{-\beta E_j-\gamma I_j}}{Z},
\end{equation}
and from \eqref{constr2} and \eqref{constr3} we have
\begin{equation*}
 \langle E\rangle
 =
 \sum_{j=1}^k E_j \frac{e^{-\beta E_j-\gamma I_j}}{Z}
 \qquad\text{and}\qquad
 \langle I\rangle
 =
 \sum_{j=1}^k I_j \frac{e^{-\beta E_j-\gamma I_j}}{Z}.
 \end{equation*}
Consequently, using the definition of the partition function \eqref{eq:partition_function}, we obtain
\begin{equation}
  \label{eq:partials_of_Z}
  -\frac{\partial\log{Z}}{\partial\beta}
  =
  \frac{\sum_{j=1}^k E_j e^{-\beta E_j-\gamma I_j}}{Z}
  =
  \langle E\rangle
  \qquad\text{and}\qquad
  -\frac {\partial\log{Z}}{\partial\gamma}
  =
  \frac{\sum_{j=1}^k I_j e^{-\beta E_j-\gamma I_j}}{Z}
  =
  \langle I\rangle.
\end{equation}
The expression \eqref{eq:entropy} for the entropy can be written as
\begin{equation}
  \label{eq:entropy2}
  S
  =
  \sum_{j=1}^k p_j(\beta E_j+\gamma I_j+\log{Z})
  =
  \beta\langle E\rangle
  +
  \gamma\langle I\rangle
  +
  \log{Z},
\end{equation}
and differentiating it with respect to $\langle E\rangle$ and using \eqref{eq:partials_of_Z} gives
\begin{equation}
  \label{eq:dSdE}
    \frac{\partial S}{\partial\langle E\rangle}
    =
    \frac{\partial\beta}{\partial\langle E\rangle}\langle E\rangle
    +
    \beta
    +
    \frac{\partial\gamma}{\partial\langle E\rangle}\langle I\rangle
    +
    \frac{\partial\log{Z}}{\partial\beta}\frac{\partial\beta}{\partial\langle E\rangle}
    +
    \frac{\partial\log{Z}}{\partial\gamma}\frac{\partial\gamma}{\partial\langle E\rangle}
    =
    \beta
    \equiv
    \frac{1}{T}.
\end{equation}
In analogy with statistical mechanics, $T=1/\beta$ is called the ``temperature'' associated with the vortex configuration, and $\beta$ is the corresponding inverse temperature, or ``coldness'' according to \citet{garrod95}. From now on we will use the term temperature in this particular sense. As can be seen from the argument above, there is, a priori, no constraint that will impose the requirement $\beta>0$ or $T>0$. That is, in general the temperature can be positive or negative, or, in fact, infinite when $\beta=0$. In particular, as $\beta$ decreases from $+\infty$ to $-\infty$, $T$ increases from $0$ through positive values to $+\infty$ (which is identified with $-\infty$), and then increases from $-\infty$ to $0$ through negative values. (One can think of this idea as traversing a circle obtained by transforming the real number line into a circle by identifying $+\infty$ and $-\infty$, both for $\beta$ and $T$; this identification is a special case of the M\"{o}bius transformation of the complex plane.) In a system with negative temperature, equation \eqref{eq:dSdE} implies that an increase in the mean energy of the system results in a decrease of its entropy; or vice versa, to increase its entropy, the system has to decrease its mean energy. Systems with negative temperatures and their behavior are discussed further in section \ref{sec:entropy-and-temperature}.

Note that the partition function \eqref{eq:partition_function} is well defined for all real values of $\beta$, including negative values, due to the finite number of terms in the sum. A similar argument as above can be used in the case of a continuum of available levels with sums replaced by corresponding integrals over the underlying phase space provided that the relevant integrals are convergent (\citet{cagliotilionsmarchioropulvirenti92,chorin,newton01}). Examples with negative temperatures will be discussed in the sections below.

We comment that additional constraints can be easily incorporated into the framework above; in that case the quantities $I$, $I_j$, and $\gamma$ can be replaced by vectors $\boldsymbol I$, $\boldsymbol I_j$, and $\boldsymbol\gamma$, respectively, and the above derivation is unchanged. By a process similar to \eqref{eq:dSdE}, we can also obtain $\partial S/\partial\langle I\rangle=\gamma$ in the case of scalar $I$ and $\gamma$, and its obvious extension in the vector case. Similarly, the second constraint can be eliminated altogether to result in a problem with just one Lagrange multiplier, $\beta$.

The model above, in which energy and additional quantities can vary, corresponds to the grand canonical case of statistical mechanics, and by rewriting \eqref{eq:entropy2} we can define (as in thermodynamics) the thermodynamic (Landau) energy potential
\begin{equation*}
  \Omega
  \equiv
  -\frac{1}{\beta}\log{Z}
  =
  \langle E\rangle
  -
  TS
  -
  \mu\langle I\rangle,
\end{equation*}
where $\mu=-\gamma T$ is usually referred to as a chemical potential. Similarly, the case with a single Lagrange multiplier corresponds to the canonical case. In this case, \eqref{eq:entropy2}, without the $\gamma$ term, can be used to define the equivalent of the Helmholtz free energy, $F$, by
\begin{equation*}
  F
  \equiv
  -\frac{1}{\beta}\log{Z}
  =
  \langle E\rangle
  -
  TS.
\end{equation*}
Consequently, additional insights can be gained from the parallels with thermodynamics (see, e.g., \citet{lim-nebus07}).

\section{The two-dimensional point vortex theory}
\label{sec:2Dvortexgas}
We now proceed to describe a two-dimensional vortex gas model (\citet{onsager49,chorin-marsden93,marchioro94,majda01}). We note that a critique of the notion of negative temperature in the two-dimensional vortex gas theory has been given by \citet{frohlich-ruelle82} and \citet{miller92}. In what follows, all vectors are expressed in the Cartesian coordinate system.

In two dimensions, vorticity is orthogonal to the plane of the flow, so we have $\boldsymbol\xi=(0,0,\zeta)$, and the vorticity equation for an isentropic, incompressible fluid flow \eqref{eq:DomegaDt} reduces to $\dfrac{D\boldsymbol\xi}{Dt}={\boldsymbol 0}$, or $\dfrac{D\zeta}{Dt}=0$. Writing the velocity in the component form, ${\bf u}=(u,v,0)$, the incompressibility condition, $\nabla\cdot{\bf u}=0$, together with the assumption that the underlying domain is simply connected implies that there exists a stream function, $\psi(x,y)$, such that
\begin{equation}
  u
  =
  \frac{\partial\psi}{\partial y},
  \qquad
  v
  =
  -\frac{\partial\psi}{\partial x}
  \label{eq:stream_velocity}
\end{equation}
and
\begin{equation}
  -\Delta\psi
  =
  \zeta.
  \label{eq:poisson_for_psi}
\end{equation}
To model a discrete collection of vortices, assume that vorticity is concentrated at points ${\bf x}_i=(x_i,y_i)$ for $i=1,\dots,n$, each with circulation $\Gamma_i$, so that
\begin{equation*}
  \zeta({\bf x})
  =
  \sum_{i=1}^n \Gamma_i \delta({\bf x-x}_i),
\end{equation*}
where $\delta$ denotes the Dirac delta function and ${\bf x}=(x,y)$. The solution to \eqref{eq:poisson_for_psi} in all of $\RR^2$ is given by
\begin{equation*}
  \psi({\bf x})
  =
  -\sum_{i=1}^n\frac{\Gamma_i}{2\pi} \log{\|{\bf x-x}_i\|},
\end{equation*}
and, using \eqref{eq:stream_velocity}, the velocity field induced by the $j$th vortex is,
\begin{equation*}
  {\bf u}_j({\bf x})
  =
  \frac{\Gamma_j}{2\pi r^2}({\bf x}-{\bf x}_j)^\perp
  =
  \frac{\Gamma_j}{2\pi r^2}\left(y_j-y,-(x_j-x)\right),
  \qquad
  r
  =
  \|{\bf x-x}_j\|.
\end{equation*}

If one assumes that each of the vortices moves under the influence of the combined velocity field of the remaining vortices, then
\begin{equation*}
  \frac{d{\bf x}_i}{dt}
  =
  \sum_{\substack{j=1\\j\ne i}}^n{\bf u}_j({\bf x}_i)
  =
  \sum_{\substack{j\ne i}}^n{\bf u}_j({\bf x}_i),
\end{equation*}
or, in the component form,
\begin{align*}
  \frac{dx_i}{dt}
  &=
  \frac{1}{2\pi}\sum_{j\ne i}\frac{\Gamma_j({y_j-y_i})}{r_{ij}^2}\\
  \frac{dy_i}{dt}
  &=
  -\frac{1}{2\pi}\sum_{j\ne i}\frac{\Gamma_j({x_j-x_i})}{r_{ij}^2},
\end{align*}
where $r_{ij}=\|{\bf x}_i-{\bf x}_j\|$. These equations form a Hamiltonian system that has rigorous connections with the Euler equation (\citet{chorin-marsden93,marchioro94}).
The corresponding Hamiltonian is
\begin{equation}
  \label{eq:hamiltonian}
  H
  =
  -\frac{1}{4\pi}\sum_{\substack{1\le i,j\le n\\i\ne j}}\Gamma_i\Gamma_j\log{\|{\bf x}_i-{\bf x}_j\|}.
\end{equation}
It is easy to check that the Hamiltonian is conserved, i.e., $\dfrac{dH}{dt}=0$, which, in particular, implies that if all the circulations are of the same sign, then the vortices cannot merge in finite time. Other conserved quantities are the total vorticity, $\Gamma$, the center of vorticity, $\bf M$, and the moment of inertia, $I$, given by 
\begin{equation}
  \label{eq:conserved_2d}
  \Gamma
  =
  \sum\Gamma_i,
  \quad
  {\bf M}
  =
  \frac{\sum\Gamma_i\bf{x}_i}{\Gamma},
  \quad
  I
  =
  \sum\Gamma_i\|{\bf x}_i-{\bf M}\|^2,
\end{equation}
where all the sums are for $i=1,\dots,n$.

We remark that on a domain with a boundary (such as a half-plane, disk, etc.), the stream function, and hence also the Hamiltonian, would be augmented by other terms to satisfy relevant boundary conditions; these terms do not significantly affect the analysis of this section. However, conserved quantities will change or be lost due to symmetry breaking in accordance with Noether's theorem (\citet{newton01}).

\begin{figure*}
  \begin{center}
  \includegraphics[height=0.29\textheight]{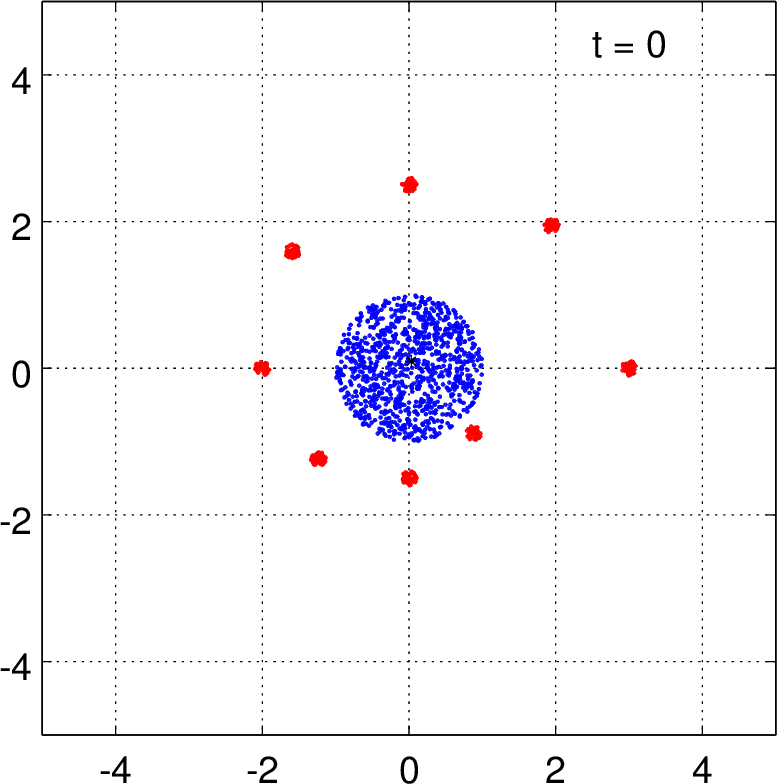}\hskip 1cm
  \includegraphics[height=0.29\textheight]{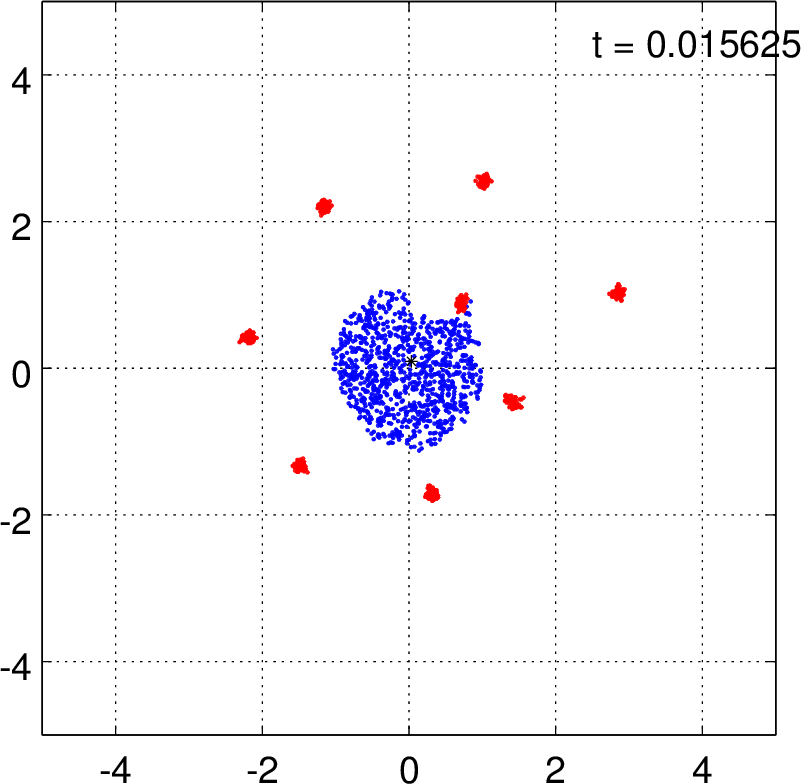}\\[\baselineskip]
  \includegraphics[height=0.29\textheight]{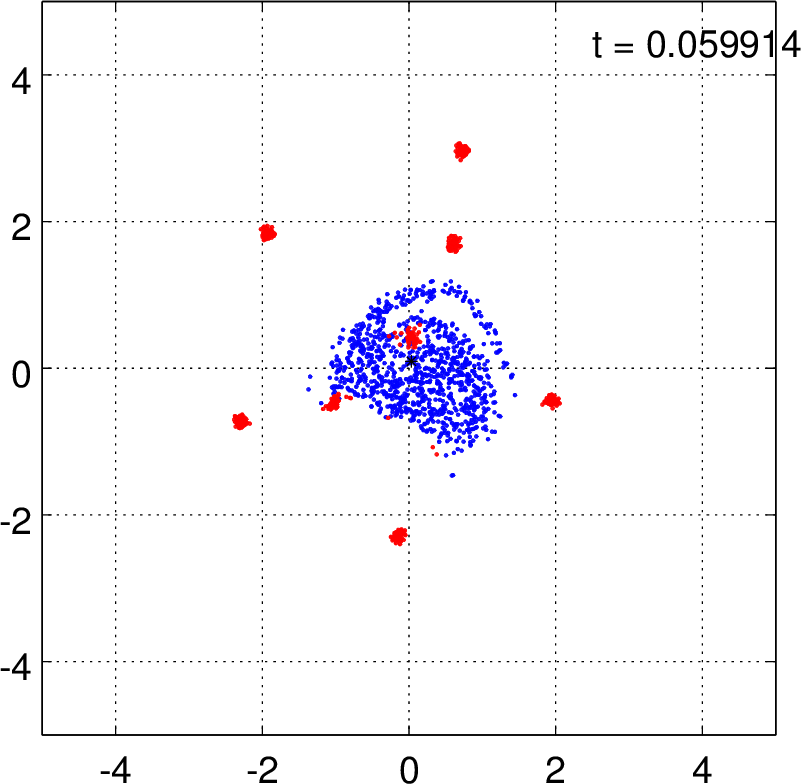}\hskip 1cm
  \includegraphics[height=0.29\textheight]{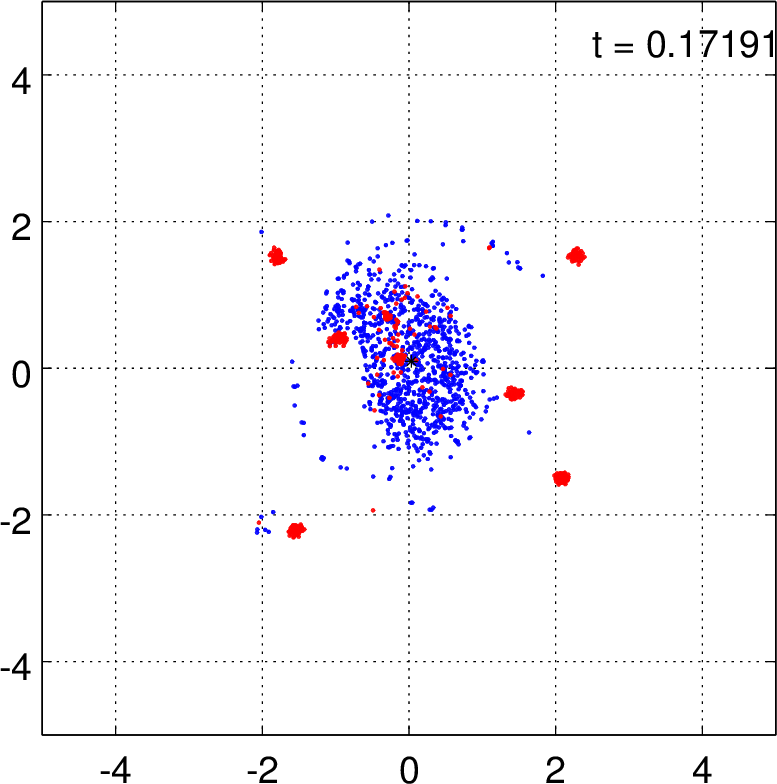}\\[\baselineskip]
  \includegraphics[height=0.29\textheight]{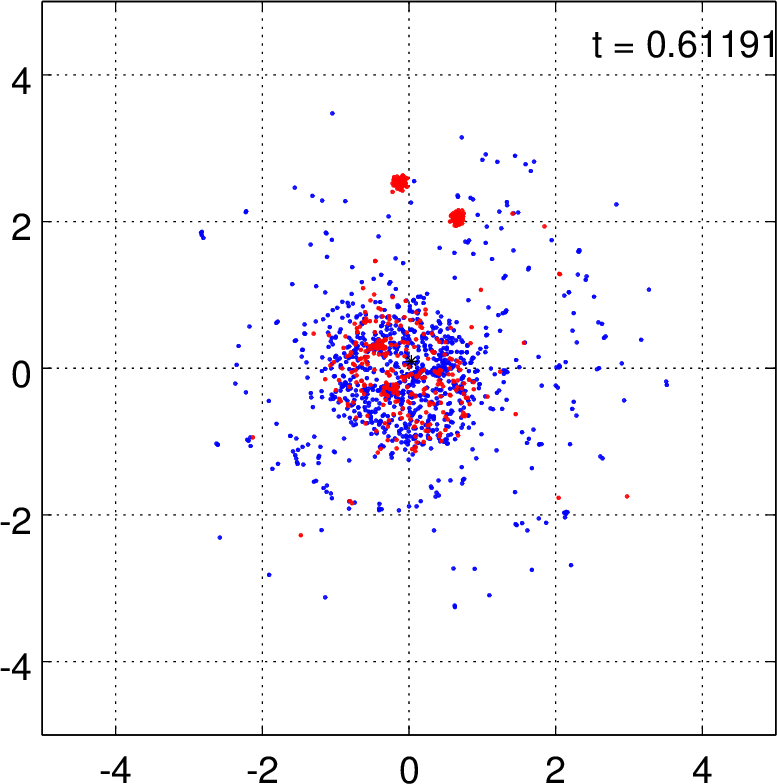}\hskip 1cm
  \includegraphics[height=0.29\textheight]{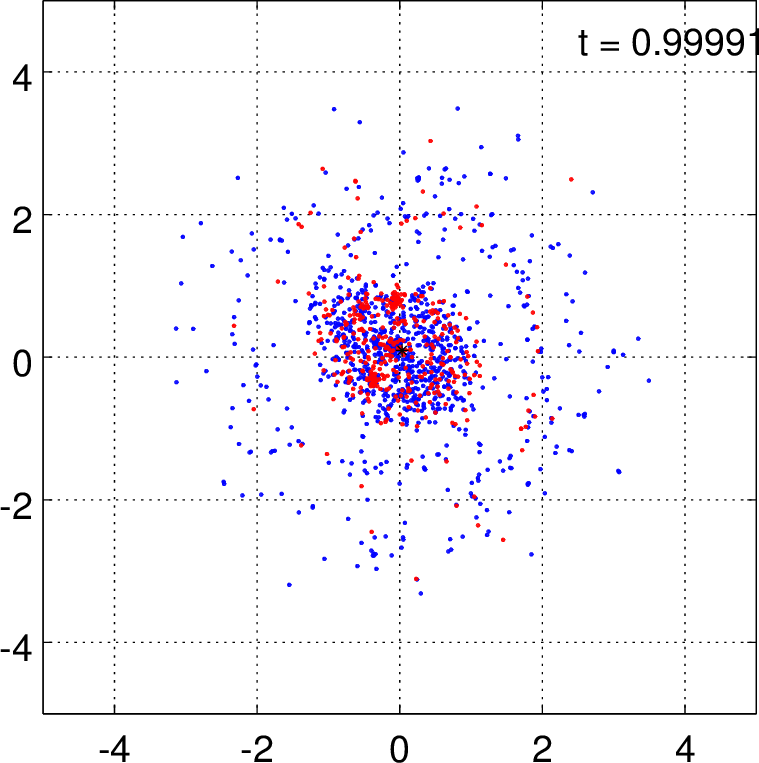}
  \end{center}
  \caption{An illustration of the dynamics of a vortex gas using point vortices in two dimensions. Smaller vortices (red) around a larger vortex (blue) are being absorbed and dispersed in the surrounding flow.}
  \label{fig:vortex_train}
\end{figure*}

Using the above theory, one can model the dynamics of vortex configurations in the plane (\citet{chorin73,marchioro94,newton01,lim-nebus07}). For example, a pair of vortices of equal circulations will rotate about the midpoint of the segment joining them with constant angular speed, while a pair of vortices of opposite circulations will translate along the line perpendicular to the segment joining them with constant speed. Also, two pairs of vortices of equal circulations will typically rotate about their center of vorticity with the vortices in each pair rotating in tandem---this can be viewed as an idealized scenario of two suction vortices rotating about one another within a larger flow, or as the rotational behavior near the ground of two arching vortices. A line of equidistant vortices of equal circulations will remain stationary, while a half-line of such vortices will roll up into a spiral, simulating a two-dimensional version of a vortex sheet roll-up. A result of a simulation with smaller vortices along a vortex sheet wrapped around a larger vortex is shown in Figure~\ref{fig:vortex_train}; both the smaller vortices, shown in red, and the larger vortex, shown in blue, are modeled by a collection of point vortices of equal strength. In section \ref{sec:entropy-and-temperature} we will briefly discuss a similar, three-dimensional scenario with a train of small, intense vortices entering a larger tornadic flow.

Aside from studying the dynamics of a vortex gas, the theory from section \ref{sec:statmech} can be applied to study the statistics of spatial distributions of vortices, originally due to \citet{onsager49}. The constraints for the Lagrange multipliers argument in the plane are given by the integral equivalent of \eqref{eq:constr1}, the Hamiltonian \eqref{eq:hamiltonian} in the canonical case, and, additionally, the conserved quantities in \eqref{eq:conserved_2d} in the grand canonical case. With a fixed number of vortices with fixed circulations $\Gamma_i$, the first of the constraints in \eqref{eq:conserved_2d} is trivially satisfied, while the second one can be satisfied for any configuration by a simple translation. This thus leads to the same scenario as discussed in section \ref{sec:statmech} with either one or two conserved quantities, the Hamiltonian (energy) and the moment of inertia.

As mentioned earlier, pairs of vortices of opposite signs of circulations will travel along a straight line with constant speed. Hence, when studying the motion of vortices in the whole plane, one typically assumes that the vortex circulations all have the same sign. The conservation of the moment of inertia given in \eqref{eq:conserved_2d} can then be invoked to conclude that the phase space of the vortex system is effectively bounded.

In the case with a bounded phase space, \citet{onsager49} shows that both positive and negative temperatures exist in the system regardless of the signs of the vortices. In particular, \citet{cagliotilionsmarchioropulvirenti92} show that in the case with $n$ vortices with equal circulations $\Gamma$ in a smooth, bounded, connected, and open domain in $\RR^2$, the integral version of the partition function \eqref{eq:partition_function} is well defined if and only if $\beta\in(-8\pi n/\Gamma_0^2,+\infty)$, where $\Gamma_0=n\Gamma$ is the total circulation of the system.

In systems with negative temperatures, the corresponding integral counterpart of \eqref{eq:p_j} implies that the most likely configurations are those with large energy $H$, which, in view of the definition \eqref{eq:hamiltonian}, corresponds to the point vortices of equal sign clustering near their center of vorticity. Thus, these ``hot'' systems would exhibit the inverse energy cascade with energy flowing from smaller scales (individual vortices) to larger scales (larger coalesced vortex). Therefore, this could serve as a simplified model for the apparent two-dimensional behavior of suction vortices moving into and within a larger tornadic flow and eventually transferring their energy into it, as discussed in the previous section.

On the other hand, in systems with positive temperatures, the most likely configurations would be those with small energy, which corresponds to vortices of equal sign spreading out. For some results of Monte Carlo simulations with positive temperatures and the moment of inertia constraint, see \citet{lim-nebus07}.

\section{Three-dimensional vortex gas models}
\label{sec:3Dvortexgas}

Modeling of three-dimensional vortex gases is much more difficult and has been developed only in special cases. Given a vorticity field ${\boldsymbol\xi}({\bf x})$ in $\mathbb{R}^3$, the system
\begin{equation*}
  {\boldsymbol\xi}
  =
  \nabla\times{\bf u},
  \qquad
  \nabla\cdot{\bf u}
  =
  0
\end{equation*}
can be solved for the velocity field $\bf u(x)$ under the assumption that ${\boldsymbol\xi}({\bf x})$ decays sufficiently fast as $\|{\bf x}\|\to\infty$ (\citet{majda01}). For the velocity field we get
\begin{equation*}
  {\bf u(x)}
  =
  -\frac{1}{4\pi}\int\frac{({\bf x-x'})\times{\boldsymbol\xi}({\bf x'})}{\|{\bf x-x'}\|^3}\,d\bf x',
\end{equation*}
known as the Biot--Savart law, and, consequently, the kinetic energy of the flow can be written as
\begin{align*}
  E
  &=
  \frac{1}{2}\int\rho({\bf x})\|{\bf u(x)}\|^2\,d{\bf x}\\
  &=
  \frac{1}{8\pi}\iint\rho({\bf x})\,\frac{{\boldsymbol\xi}({\bf x})\cdot{\boldsymbol\xi}({\bf x'})}{\|{\bf x-x'}\|}\,d{\bf x'}\,d{\bf x}
  +
  \frac{1}{8\pi}\iint\frac{(\nabla\rho({\bf x})\times{\bf u(x)})\cdot{\boldsymbol\xi}({\bf x'})}{\|{\bf x-x'}\|}\,d{\bf x'}\,d{\bf x},
\end{align*}
which, in the case of a homogeneous fluid rescaled so that $\rho({\bf x})\equiv1$, reduces to (\citet{chorin})
\begin{equation}
  \label{eq:E}
  E
  =
  \frac{1}{8\pi}\iint\frac{{\boldsymbol\xi}({\bf x})\cdot{\boldsymbol\xi}({\bf x'})}{\|{\bf x-x'}\|}\,d{\bf x'}\,d{\bf x}.
\end{equation}
All of the above integrals are taken over the supports in $\RR^3$ of the relevant functions. Expressions with a kinetic energy of this or similar form appear in most of the three-dimensional vortex gas models. In the following sections we first briefly mention the models of \citet{lionsmajda00} and \citet{berdichevsky98,berdichevsky02}, which do not directly apply in our context as they deal with nearly parallel vortices. We then focus more on the model of \citet{chorin-akao91}, also described in \citet{chorin}, which deals with the effects of stretching and folding of vortices, and we also briefly mention some consequences of a follow up work by \citet{flandoligubinelli02}, which shows that vortices with fractal cross sections have finite energy.

\subsection{Models with Nearly Parallel Vortices}
Extending the work of \citet{kleinmajdadamodaran95}, \citet{lionsmajda00} develop a mathematically rigorous equilibrium statistical mechanics theory for a collection of three-dimensional, periodic, nearly parallel interacting vortices with equal circulations, taken equal to $1$. In this theory, simplified asymptotic expansions of the Navier--Stokes equations with a large Reynolds number, derived by \citet{kleinmajdadamodaran95}, are utilized. The vortex model is described by the following Hamiltonian equations of motion of each vortex, $X_j(\sigma,t)\in\Real^2$, $j=1,\dots,N$,

\begin{equation*}
  \frac{\partial X_j}{\partial t}
  =
  J
  \left[
    \alpha
    \frac{\partial^2 X_j}{\partial\sigma^2}
    +
    \frac{1}{2}\sum_{k\ne j}^N
      \frac{X_j-X_k}{\|X_j-X_k\|^2}
  \right],
\end{equation*}
where $t$ denotes time, $\sigma$ parametrizes the center curve of each vortex, $\alpha>0$ is a parameter related to the vortex core structure and taken to be the same for all vortices, and $J=\begin{bmatrix}0 & -1 \\ 1 & 0\end{bmatrix}$. The first term in the brackets models the self-interaction behavior of the $j$th vortex, while the second term models the motion of the $j$th vortex due to the other vortices. Note that all of these interactions are restricted to the same fixed height $\sigma$. This model, therefore, reduces to the two-dimensional model considered in section \ref{sec:2Dvortexgas} if the dependence on $\sigma$ is suppressed. The main result is the derivation of a mean-field theory for the presented model. To this end, a limit as $N\to\infty$ of a rescaled model is studied and a probability distribution of vortex filaments in $\Real^2$ is obtained. This model allows for a limited amount of stretching and folding, inhibiting the transfer of energy across scales. Also, only positive temperatures are considered.

Another statistical mechanics theory for periodic vortex lines, this time in a cylindrical bounded domain and under the assumption that the underlying Hamiltonian system is ergodic, is developed by \citet{berdichevsky98,berdichevsky02}. The model uses the maximum entropy principle and an assumption of sufficiently smooth vortex lines; the smoothness is controlled by a parameter referred to as ``vortex diffusivity.'' In this approach, the kinetic energy of the system is determined as a solution to a variational problem, and the partitioning of the corresponding phase space with respect to this energy is analyzed via a maximization of an entropy functional. A comparison to the result of \citet{lionsmajda00} is given, which includes a mild criticism of the latter and pointing out that Berdichevsky's model allows for negative temperatures. Nevertheless, the amount of stretching and folding of vortices is limited in the same way as in \citet{lionsmajda00}.

\subsection{Models with Folding Vortices}
In contrast to the models in the previous section, the main feature of Chorin's model (\citet{chorin-akao91,chorin91,chorin}) is folding of vortex filaments leading to the transfer of energy from smaller to larger scales. The model is described in the context of a single vortex filament on a (cubic) lattice, $\ZZ^3$, although it can naturally be applied to a collection of such vortices. The vortices are identified with oriented, self-avoiding random walks on the lattice made up of vertical and horizontal line segments connecting adjacent points on the lattice.

The expression \eqref{eq:E} for the kinetic energy of one vortex is discretized and becomes
\begin{equation}
  \label{eq:E_discrete}
  E
  =
  \frac{1}{8\pi}\sum_i\sum_{j\ne i}\frac{{\boldsymbol\xi_i}\cdot{\boldsymbol\xi_j}}{\|i-j\|}
  +
  \frac{1}{8\pi}\sum_ iE_{ii},
\end{equation}
where $i$ and $j$ are the three-dimensional coordinates of the locations of the centers of the vortex segments making up the vortex, $E_{ii}$ is the ``self-energy'' term of the $i$th segment that is constant and is, therefore, neglected in the analysis, $\|i-j\|$ is the distance between the $i$th and $j$th segment centers, and $\boldsymbol\xi_i$ is the vorticity of the $i$th segment. Notice that after neglecting the self-energy term, there is no singularity in \eqref{eq:E_discrete}, although a logarithmic singularity will appear if the lattice spacing is allowed to approach $0$.

The probability of a vortex with energy $E$ and inverse temperature $\beta$ is given by $P(E)=e^{-{\beta E}}/Z$, where $Z$ is a corresponding partition function. The phase space of permissible configurations with $N$ segments is then explored and analyzed using Monte Carlo techniques coupled with a Metropolis--Hastings rejection algorithm. A small interval of positive, negative, and zero values of $\beta$ is explored; equilibrium configurations with negative temperatures are shown to be straighter than those with positive temperatures, which tend to kink up and fold into balls. The $\beta=0$ case (infinite temperature) corresponds to the well-studied polymeric case (see, e.g., \cite{chorin}).

The effects of stretching are studied by performing simulations with vortices of various lengths (but fixed lattice spacing). One observation is that the mean energy of a vortex filament with a fixed temperature increases with the length. Also, if the mean energy is constant and a vortex is being stretched, the temperature decreases from negative to positive through $|T|=\infty$ ($\beta=0$). Therefore, since the energy of the flow should be conserved, if a vortex with negative temperature is being stretched, it has to ``cool off'' and/or folding has to occur to cancel any excess energy. Furthermore, when the temperature decreases to the positive range, folding should occur as discussed in the previous paragraph. Finally, it is argued that the entropy of a vortex increases with its length.

\citet{flandoligubinelli02}, motivated by the work of Chorin and some numerical experiments, consider single three-dimensional vortices with Brownian or smooth cores, possibly fractal cross sections, and positive and negative temperatures. This approach thus alleviates the constraints of Chorin's lattice model and incorporates Chorin's argument for fractal cross section (\citet{chorin}). The authors show that vortices with cross sections of fractal dimension greater than $1$ can exist with finite energy and finite partition functions over all positive temperatures and a limited range of negative temperatures. Folding of a vortex is naturally present in the model, and so features present in Chorin's model are present in this model as well.

We conclude this section by commenting on the appropriateness of this model and repeating some of the points made earlier in this paper. As mentioned earlier, visual observations of small-scale subvortices in a tornado show intense vortices with transient life spans that start off smooth and straight, then kink up and dissipate. These are the supercritical or suction vortices, and tracks left by such vortices generally are associated with considerable damage indicating high energy density. This mirrors the modeled behavior of the negative-temperature vortices undergoing stretching which kink up and then dissipate. We also point out that \citet{lewellensheng80} modeled tornadic flows using a large-eddy simulation model, where the large eddies could be identified with supercritical vortices. In this paper, we draw the connection between large eddies and negative-temperature vortices that transfer energy to the larger scale flow, thus driving the turbulence.

In the next section we discuss the consequences and importance of negative-temperature systems and the effects of stretching and folding of vortices.

\section{Entropy and temperature}
\label{sec:entropy-and-temperature}
In classical thermodynamics, when two isolated systems in equilibrium are brought into thermal contact, heat will flow from the ``hotter,'' or higher-temperature system, to the ``colder,'' or lower-temperature one. In this section, we use this idea to show that in this sense negative temperatures are hotter than positive temperatures and also negative temperatures that are closer to zero are hotter than negative temperatures that are farther from zero. We follow the treatment given in \citet{landaulifshitz}.

Consider two isolated systems, one with mean energy $\langle E_1\rangle$, entropy $S_1$, and temperature $T_1$ and the other with mean energy $\langle E_2\rangle$, entropy $S_2$, and temperature $T_2 \ne T_1$, each separately in equilibrium. Assume that the two systems are brought into thermal contact so that the mean energy, $\langle E\rangle$, and the entropy, $S$, of the combined system are
\begin{equation*}
  \langle E \rangle
  =
  \langle E_1\rangle
  +
  \langle E_2\rangle
  \qquad\text{and}\qquad
  S
  =
  S_1+S_2.
\end{equation*}
As the combined system adjusts to equilibrium, the time rate of change of the total entropy is positive and satisfies
\begin{equation*}
  \frac{dS}{dt}
  =
  \frac{dS_1}{dt}
  +
  \frac{dS_2}{dt}
  =
  \frac{dS_1}{d\langle E_1\rangle}\frac{d\langle E_1\rangle}{dt}
  +
  \frac{dS_2}{d\langle E_2\rangle}\frac{d\langle E_2\rangle}{dt}
  >
  0.
\end{equation*}
Conservation of energy implies that
\begin{equation*}
  \frac{d\langle E\rangle}{dt}
  =
  \frac{d\langle E_1\rangle}{dt}
  +
  \frac{d\langle E_2\rangle}{dt}
  =
  0,
\end{equation*}
and hence, with  $\beta_i=1/T_i$,
\begin{equation*}
  \frac{dS}{dt}
  =
  \left[
    \frac{dS_1}{d\langle E_1\rangle}
    -
    \frac{dS_2}{d\langle E_2\rangle}
  \right]
  \frac{d\langle E_1\rangle}{dt}
  =
  \left(
    \beta_1
    -
    \beta_2
  \right)
  \frac{d\langle E_1\rangle}{dt}
  >
  0,
\end{equation*}

It follows that when $\beta_1>\beta_2$, heat will flow from system 2 to system 1 (since then $d\langle E_1\rangle/dt>0$ and $d\langle E_2\rangle/dt<0$), so system 2 is hotter than system 1. In particular, if $\beta_1$ and $\beta_2$ have the same sign, we have $T_1<T_2$, whether positive or negative, which is consistent with the natural interpretation of larger temperatures corresponding to hotter systems. Notice, however, that if $\beta_1$ and $\beta_2$ have opposite signs, we have $0<T_1<\infty$ and $-\infty<T_2<0$, which shows that a negative-temperature system is hotter than a positive-temperature system.
Hence, the three cases below follow: if
\begin{enumerate}
  \item $0<T_1<T_2 \leq \infty$, or
  \item $-\infty \leq T_2<0<T_1  \leq \infty$, or
  \item $-\infty  \leq T_1<T_2  < 0$,
\end{enumerate}
then the system with temperature $T_2$ is hotter than the system with temperature $T_1$ and energy will be transferred from the former to the latter.

We now consider a collection of vortices (a vortex gas), with a specific example of interest being a larger tornadic vortex with ambient smaller vortices. The entropy of such a system is affected by the structure of each vortex, e.g., their stretching, folding, and collapse, and also by the spatial organization, or configuration, of the vortices.

In a system with a single vortex, the entropy is a function of the structure of the vortex, and we can refer to it as a structural entropy of the vortex. In the following paragraphs we will focus on briefly discussing supercritical vortices and argue that they have negative temperature. We first focus on relevant previous work on vortices in a Ward vortex chamber and then discuss connections to supercritical vortices in nature, such as the suction vortices discussed earlier.

Experimental vortex chamber studies by \citet{ward72} and later \citet{church77} have resulted in the following observations. By controlling the volumetric flow of the updraft and the angular momentum added to the flow (or swirl), it has been observed that when the ratio of the two is in a certain range, then the flow configuration takes on a structure that resembles a side view of a champagne glass: the stem of the glass corresponds to the supercritical (what we will argue to be a negative-temperature) vortex, and the part of the glass above the stem corresponds to the breakdown bubble (below the subcritical vortex). In particular, for swirl ratios in the relevant range, keeping the updraft in the convection chamber fixed and increasing the angular momentum in the convergence zone of the chamber resulted in decrease of the radius of the supercritical vortex and its length, while the axial velocity as well as the azimuthal velocity of the supercritical vortex increased.

\citet{fiedlerrotunno86} combined the results of these experimental vortex chamber studies and the theoretical studies by \citet{barcilon67,burggrafstewartsonbelcher71,burggraf77}, all of which contributed to the identification of quantitative relationships between the radius of the supercritical vortex and the amount of angular momentum added to the flow, as well as relationships between its axial and azimuthal velocity components and the added angular momentum.

In the models considered in \citet{barcilon67,burggrafstewartsonbelcher71,burggraf77}, the effect of the boundary layer on the structure of supercritical vortices is studied. In these models, the boundary layer is decomposed into a primary layer immediately above the ground and a secondary layer above that. The surface friction in the primary boundary layer kills off the cyclostrophic balance, and the pressure gradient force dominates. The result is a nearly radial inflow that erupts from the surface to form a vortex. In the secondary boundary layer, the flow recovers to a potential flow, and above the secondary boundary layer the cyclostrophic balance is recovered.

We now argue that such supercritical vortices have negative temperature in the statistical mechanics sense. To this end, we first observe that, based on the arguments above, when the supplied angular momentum is increased, the axial and azimuthal components of the velocity increase (while the radial one is negligible), so the (kinetic) energy density in the vortex increases. We next argue that the entropy density has to be decreasing in this case, resulting in the negative temperature in the system.

Intuitively, since the radius and the length of the vortex are decreasing with increasing angular momentum, the volume of the vortex is decreasing, and the vortex flow is becoming more organized, resulting in a decrease in entropy density.

More specifically, we can argue using considerations about the corresponding phase space. Entropy can alternatively be defined as the logarithm of the measure of the subset of configurations in the phase space with the given energy. These configurations of a single vortex would be described by the geometry of the vortex (such as the center line and the cross section) as well as its vorticity. As discussed before, vortices with very large energies would be straight and narrow, with high values of vorticity. As such, the set of such configurations would have a small measure and thus small entropy. Increasing energy would then result in further decrease of entropy, and thus such configurations would have negative temperatures. Applying this argument locally and assuming local equilibrium, it follows that a supercritical vortex has negative temperature.

As in the paper by \citet{fiedlerrotunno86}, who argue that tornadoes exhibit similarities to vortices in vortex chambers, we believe that under some conditions supercritical vortices in nature (suction vortices) also have negative temperatures. The crucial fact here is that the vortices must be in contact with the ground which impedes the cyclostrophic balance; in this case, a narrow, intense, negative-temperature vortex can form if the swirl ratio is in the right range. The effects of the boundary layer are important in that as the angular momentum increases, the thickness of the boundary layer decreases, and the radius of the supercritical vortex (thought of as an extension of the boundary layer) also decreases, as does the length of the vortex. Also, both the vertical velocity in the core and the azimuthal velocity increase. We emphasize here that the swirl ratio must be in a certain range for the vortex to be supercritical; for a swirl ratio outside this range the vortex behavior would be different, and the temperature would not necessarily be negative. Discussions of relevant ranges for swirl ratios can be found in \citet{bluestein13,churchsnowbakeragee79}. Another mechanism that may lead to decrease in entropy density and increase in energy density is a corner flow collapse as discussed in \citet{lewellens07a,bluestein13}. In this case, it is thought that the rear flank downdraft blocks the radially inward-flowing air, thus blocking the influx of angular momentum. This would lead to an intensification of the near-surface vortex and the possible creation of a supercritical vortex.

Consider now two disjoint vortex systems: one consisting of an intense, supercritical vortex and one consisting of a larger, turbulent one (tornado or a developing tornado; turbulence would correspond to $T=\pm\infty$). Assume that the two systems are each separately in equilibrium, and the supercritical vortex moves into the tornado. Since the supercritical vortex has negative temperature, presumably higher than the tornadic vortex, the discussion from the beginning of the section implies that the supercritical vortex will lose energy to the larger tornadic flow, thus contributing to the maintenance (or genesis) of the tornado. The supercritical vortex can either fold and dissipate or breakdown into multiple smaller subvortices. Videos of tornadoes appear to show both possibilities occurring.

\begin{figure*}
  \begin{center}
  \includegraphics[width=3.6in]{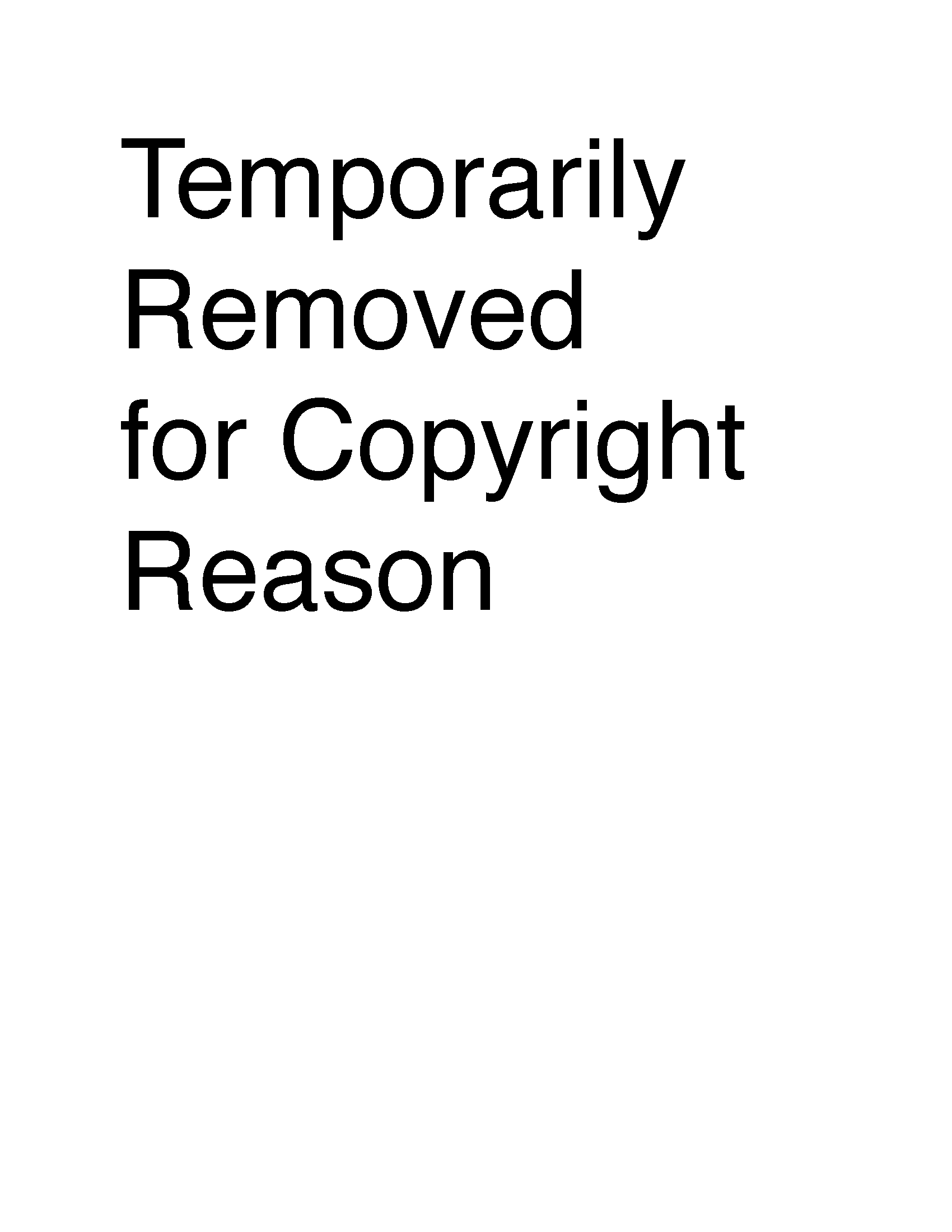}
  \end{center}
  \caption{Vortex sheet roll-up over a lake with traces of vortices visible on the water surface. \copyright~Gene Moore}
  \label{fig:Vortex sheet roll-up along the rear flank gust front}
\end{figure*}

Applying this argument to a vortex gas scenario with multiple supercritical vortices entering a tornado (such as those generated in a vortex sheet and shown in Figure \ref{fig:Vortex sheet roll-up along the rear flank gust front} or those observed in the simulations of \citet{orfwilhelmsonwickerleefinley14,orf16}), the above process repeats itself as the individual vortices dissipate and transfer energy to the surrounding flow, thus contributing to the maintenance of the tornado or genesis of the developing tornado and exhibiting an inverse energy cascade as in the case of point vortices in two dimensions.

\begin{figure*}
  \begin{center}
    \includegraphics[width=\textwidth]{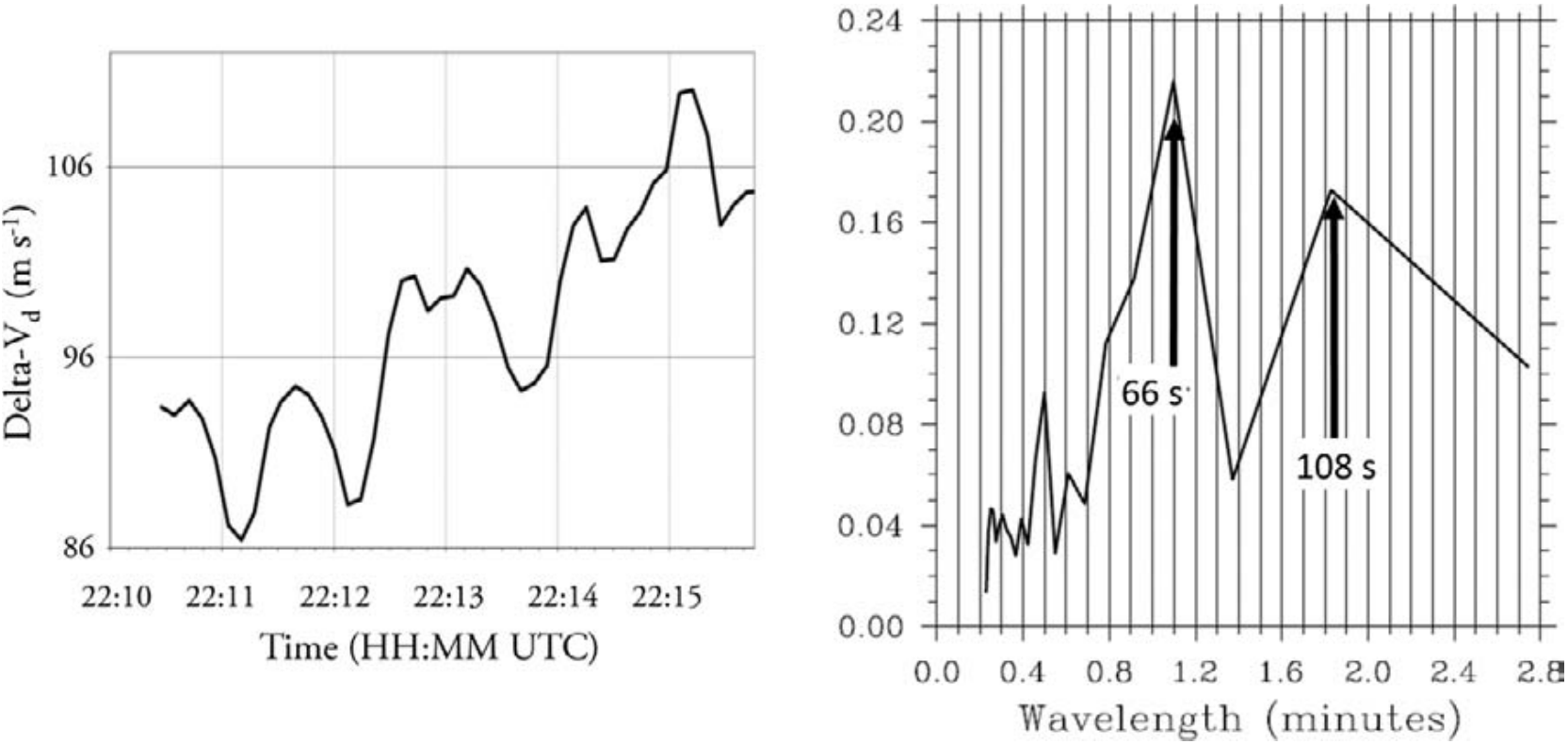}
  \end{center}
  \caption{Time series of maximum gate-to-gate shear, $\Delta V$, consistent with energy being pumped into the tornado in discrete pulses, possibly from roll-up vortices within a vortex sheet (left); FFT of $\Delta V$ with peak energy at $66$ seconds and $108$ seconds (right), suggestive of rotating asymmetry in the vortex; \copyright~AMS, \citet{wurman2013}.}
  \label{fig:wurman-kosiba-fig7}
\end{figure*}

A train of vortices in a vortex sheet entering the tornado and transferring the energy to the larger scale is a possible explanation for the gate-to-gate shear measured for the Goshen County, Wyoming tornado and displayed in Figure \ref{fig:wurman-kosiba-fig7} (\citet{wurman2013}). In this figure we observe periodic pulses in the gate-to-gate shear which could possibly be explained by intense vortices entering the tornado. It should be pointed out that no multiple vortices were visually observed for this tornado (\citet{wurman2013}), but the existence of the small scale supercritical vortices cannot be ruled out, since they would be very hard to observe both visually and on a radar.

\section{Conclusions}
\label{sec:conclusions}
In this work we have summarized some of the main features of vortex gas theories in two and three dimensions that appear to be relevant to the genesis and maintenance of a tornado. To this end, the statistical mechanics aspects of these theories have been reviewed in section \ref{sec:statmech}, the two-dimensional point vortex theory was presented in section \ref{sec:2Dvortexgas}, and the most relevant results of three-dimensional vortex gas models have been summarized in section \ref{sec:3Dvortexgas}. The statistical thermodynamic aspects of interactions of high-temperature vortices (represented by small-scale intense suction vortices) with cooler ambient vortices (represented by a larger tornadic vortex) have been discussed in section \ref{sec:entropy-and-temperature}.

One of the main results of this work is the proposed explanation of the inverse energy cascade mechanism using the vortex gas theory, which can be used to explain tornadogenesis and maintenance. Using the vortex gas theory we relate supercritical vortices to vortices with negative temperature and show that dissipation or breakdown of supercritical vortices results in the transfer of energy from smaller to larger scales. Indeed, video footage of subvortices in tornadoes suggests that they behave as negative-temperature vortices would. For example, in some instances the vortices' appearance is associated with stretching and with strong convergence. This may indicate that the vortex intensification is related to a decrease in entropy and an increase in energy. Numerical simulations of intense vortices and their interpretations in \citet{fiedlerrotunno86,fiedler94,lewellensxia00,xialewellens03,lewellens07,lewellens07a} show that the maximum wind speeds in intense narrow vortices undergoing vortex breakdown may exceed the speed of sound in the vertical direction. The suction vortices in nature, which may be extremely intense, have been observed to pull cornstalks out of clay soil with their roots.

In this work we mainly consider barotropic vortices, but this context is viewed as subsequent to preliminary stages of vertical vorticity production from horizontal vorticity, for which possible mechanisms of production include baroclinic production in the rear flank downdraft (\citet{markowskirichardsonbryan14}). We also highlight the importance of the boundary layer and its relation to the structure of the supercritical suction vortices.

\begin{figure*}
  \begin{center}
    \includegraphics[width=\textwidth]{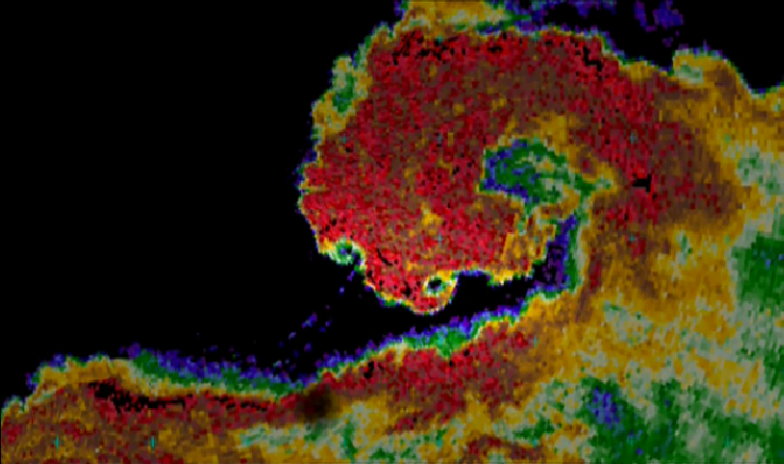}
  \end{center}
  \caption{A reflectivity image of a tornado showing potential subvortices on the periphery of a larger tornado; \copyright~Joshua Wurman, \citet{nova04}.}
  \label{fig:tornado_fractal}
\end{figure*}

As we noted earlier in section \ref{sec:vortexstretching}, under the influence of strong rotation, a turbulent flow becomes anisotropic with the flow tending toward, but never fully becoming, a two-dimensional flow (\citet{pouquet10}). Also, most of the relevant three-dimensional theories, with the exception of Chorin's, deal with nearly parallel vortices, and thus we believe that valuable insight can be gained from the two-dimensional theory as well. In particular, the tracks of overlapping suction vortices moving through fields shown in Figure \ref{fig:suctionspots} as observed by \citet{fujita81} and others (\citet{grazulis97}) from the air can be modeled by a pair of cyclonically rotating point vortices in the half-plane; also, the two-dimensional behavior of interacting vortices of various strengths can be modeled (see Figure \ref{fig:vortex_train}) and exhibits features similar to the dissipation of a smaller, intense vortex in a larger tornadic flow, even though this model cannot capture any three-dimensional effects. We note that it is not unreasonable to expect a scenario similar to that shown in Figure \ref{fig:vortex_train} in nature, as the radar reflectivity image shown in Figure \ref{fig:tornado_fractal} indicates.
In this image, a hook echo region of a supercell thunderstorm is shown, with additional, smaller hooks on the boundary of the region likely representing successive vortices in a vortex sheet. Such vortices could provide periodic pulses of energy to the tornado in accordance with the main ideas of this paper.

\bibliographystyle{abbrvnat}
\bibliography{vortex}

\end{document}